# High quality $Al_{0.37}In_{0.63}N$ layers grown at low temperature (<300°C) by radio-frequency sputtering

A Núñez-Cascajero[1,2], R. Blasco[1], S Valdueza-Felip[1], D. Montero[3], J. Olea[3] and F. B. Naranjo[1]

[1]GRIFO, Departamento de Electrónica, Universidad de Alcalá, 28871 Alcalá de Henares, Spain
[2]Departamento de Ingeniería Mecánica, Universidad Carlos III de Madrid, 28911 Leganés, Spain
[3]Departamento de. Estructura de la Materia, Física Térmica y Electrónica, Universidad Complutense de Madrid, 28040 Madrid, Spain

E-mail: arantzazu.nunez@uah.es

High-quality $Al_{0.37}In_{0.63}N$ layers have been grown by reactive radio-frequency (RF) sputtering on sapphire, glass and Si (111) at low substrate temperature (from room temperature to 300°C). Their structural, chemical and optical properties are investigated as a function of the growth temperature and type of substrate. X-ray diffraction measurements reveal that all samples have a wurtzite crystallographic structure oriented with the *c*-axis perpendicular to the substrate surface, without parasitic orientations. The layers preserve their Al content at 37 % for the whole range of studied growth temperature. The samples grown at low temperatures (RT and 100°C) are almost fully relaxed, showing a closely-packed columnar-like morphology with an RMS surface roughness below 3 nm. The optical band gap energy estimated for layers grown at RT and 100°C on sapphire and glass substrates is of ~2.4 eV while it red shifts to ~2.03 eV at 300°C. The feasibility of growing high crystalline quality AlInN at low growth temperature even on amorphous substrates open new application fields for this material like surface plasmon resonance sensors developed directly on optical fibers and other applications where temperature is a handicap and the material cannot be heated.

## 1. Introduction

$Al_xIn_{1-x}N$ is an interesting III-nitride ternary alloy because it presents a wide direct band gap which can be tuned from the near infrared (InN: 0.7 eV [1]) to the deep ultraviolet (AlN: 6.2 eV [2]). It has application in electronics, optoelectronics and sensors fields. $Al_xIn_{1-x}N$ with a composition of $x$=0.83 grows lattice matched to GaN [3] increasing the applications of the material in high electron mobility transistors [4,5] and Bragg reflectors [6,7]. Other applications of $Al_xIn_{1-x}N$ material is as active layer in multijunction solar cells [8–10]. In spite of the potential properties and applications of $Al_xIn_{1-x}N$, it is the least studied alloy among III-nitrides due to the difficulty of growing high-quality and single-phase layers because of the large differences between its binaries (InN and AlN) constituents: *i.e.* the differences in bonding energies [11], lattice mismatch and growth temperature [12] are responsible of a large immiscibility gap [13,14], and thus for the phase separation and composition inhomogeneities commonly present in the alloy.

The growth of $Al_xIn_{1-x}N$ layers has been reported by different growth techniques such as metal-organic chemical vapor deposition (MOCVD) [11,15–18], molecular beam epitaxy (MBE) [19–23] and sputtering deposition [10,24–40]. The first two growth techniques allow the achievement of high crystal quality layers at high growth temperatures; but the use of the sputtering low cost technique, allows the deposition of polycrystalline layers at low temperatures, enabling the growth even on flexible substrates, thanks to the high kinetic energy of the ions involved in the growth process. The growth of $Al_xIn_{1-x}N$ at low temperatures (< 300°C) by sputtering has been reported by several groups [28–32,35,40] on different substrates. Most of them use a mixture of Ar and $N_2$ atmosphere which leads to higher deposition rates at the expense of layer crystal quality as previously discussed in [41,42]. In particular, Afzal *et al.* studied the influence of the substrate temperature (from RT to 300°C) on the properties of $Al_xIn_{1-x}N$ layers grown on Si (111) under a mixture of Ar and $N_2$ plasma [28]. It should be remarked that for the $Al_xIn_{1-x}N$ grown at RT they do not achieve crystalline $Al_xIn_{1-x}N$ layers. He *et al.* studied the effect of the growth temperature (from RT to 500°C) on $Al_xIn_{1-x}N$ grown on quartz by sputtering, showing the appearance of (103) orientation for growth temperatures below 300°C [35]. Literature reveals that AlInN growth is highly influenced by the substrate type and growth conditions.

In this work, we demonstrate the successful growth of well oriented single phase crystalline $Al_xIn_{1-x}N$ layers with x~0.37 on sapphire, silicon (111) and glass substrates from RT to 300°C by RF-sputtering under a pure nitrogen atmosphere. The effect of the type of substrate and growth temperature on the structural, chemical, morphological and optical properties of the grown layers is deeply analyzed.

## 2. Experimental methods

The (0001)-oriented sapphire, glass and p-Si (111) substrates were chemically cleaned in organic solvents and blown dry with nitrogen before being loaded in the sputtering chamber. An AJA International, ATC ORION-3-HV reactive radiofrequency (RF) sputtering system is used for growing the $Al_xIn_{1-x}N$ samples. The targets, pure In (99.995%) and pure Al (99.999%), are pre-sputtered with Ar (99.9999%) prior to the growth. Then the substrates were loaded into the growth chamber, where they were outgassed for 30 min 100°C above the growth temperature. A surface pretreatment with Ar was then applied to increase the substrate surface cleaning. After this procedure, the base pressure achieved after cooling down the substrate to the growth temperature was ~$10^{-6}$ Pa. The $Al_xIn_{1-x}N$ films are grown by co-sputtering of In and Al targets, mounted in separated magnetron guns, in a pure N (99.9999%) ambient. The substrate-target distance, nitrogen flow, sputtering pressure and RF power applied to the indium and aluminum targets were kept at 10.5 cm, 14 sccm, 0.47 Pa, 40 W and 150 W, respectively. A sputtering deposition time of 4 h was used for all the samples. A thermocouple placed in direct contact with the substrate holder is used to monitor the substrate temperature ($T_s$) during deposition. It was ranged from room temperature (RT) to 300°C (see Table 1). Further details of the growth procedure of the layers are reported in previous works from our group [26,27].

**Table 1:** Summary of the samples under study classified as a function of substrate and growth temperature. Aluminum content (*x*) extracted from HRXRD measurements for samples S1-S6 and from WDX for S7 and S8, FWHM of the rocking curve of the (0002) $Al_xIn_{1-x}N$ diffraction peak, RMS surface

roughness calculated from AFM micrographs, layer thickness estimated from FESEM images, optical properties estimated from transmission measurements.

| Sample | Substrate | $T_s$ (ºC) | Al content x | FWHM rocking curve (°) | RMS (nm) | Thickness (nm) | Growth rate (nm/h) | $E_g$ (eV) | ΔE (meV) |
|---|---|---|---|---|---|---|---|---|---|
| S1 | Sapphire | RT | 0.37 | 6.5 | 2.5 | 510 | 255 | 2.33 | 288 |
| S2 | Si (111) | | 0.37 | 6.4 | 1.6 | 485 | 243 | - | - |
| S3 | Glass | | 0.32 | 5.8 | 2.0 | - | | 2.37 | 345 |
| S4 | Sapphire | 100 | 0.36 | 6.0 | 2.7 | 345 | 173 | 2.44 | 289 |
| S5 | Si (111) | | 0.39 | 6.7 | 2.9 | 446 | 223 | - | - |
| S6 | Glass | | 0.34 | 5.2 | 2.4 | - | | 2.46 | 388 |
| S7 | Sapphire | 300 | 0.36 | 1.4 | 5.8 | 530 | 265 | 2.03 | 154 |
| S8 | Si (111) | | 0.37 | 3.3 | 3.1 | 840 | 420 | - | - |

High-resolution X-ray diffraction (HRXRD) measurements (using a PANalytical X'Pert Pro MRD system) were used for estimating the crystalline orientation and composition of the deposited films. The surface morphology was assessed by atomic force microscopy (AFM) using a Veeco Dimension 3100 microscope in tapping mode. The data processing and image generation was done with WSxM software [43]. The cross-section and surface morphology of the layers were characterized with a Zeiss Ultra 55 field-emission scanning electron microscope (FESEM). The samples were electrically characterized by means of sheet resistance and Hall effect measurements with the Van der Pauw configuration at room temperature using a Keithley SCS 4200 model with four Source and Measure Units. Samples were 1×1 cm$^2$ pieces with four In contacts in the corners. The magnetic field used in Hall effect measurements was 0.88 T and the injected current was kept fixed at 10 µA. For each configuration the polarity of the current source and the direction of the magnetic field were changed, (for a total of 16 measurements), in order to minimize spurious thermos-galvanomagnetic effects. Finally, transmittance measurements of the $Al_xIn_{1-x}N$ on sapphire and glass layers performed at normal incidence in the 350–1700 nm wavelength range using an optical spectrum analyzer were used to estimate their optical properties.

## 3. Results and discussion

*3.1. Structural and morphological characterization*

**Figure 1** shows a comparison of the HRXRD patterns of the $Al_xIn_{1-x}N$ films deposited on different substrates at RT and 100°C. The HRXRD results of the $Al_xIn_{1-x}N$ layers grown at 300°C on sapphire and silicon were presented in [27] and [26], respectively. All the layers under study present crystalline wurtzite structure oriented with the *c*-axis perpendicular to the substrate surface. The diffraction peak related to the substrate (0006) and (111) for sapphire and Si, respectively is also present in the corresponding pattern. In the case of the $Al_xIn_{1-x}N$ layer grown at 100°C on silicon, HRXRD measurements show a peak at ~60° which corresponds to the silicon (222) diffraction peak. From the $Al_xIn_{1-x}N$ layer only the (0002) and (0004) diffraction peaks are detected. These results confirm the capability of the low cost sputtering technique for the development of crystalline III-nitride layers with wurtzite structure even at RT. The review of the literature shows that usually $Al_xIn_{1-x}N$ layers deposited at low temperature are not single phase crystalline, showing parasitic reflections. In particular Afzal *et al.* did not obtain crystalline $Al_xIn_{1-x}N$ layers on Si (111) when growing at RT [28]. And He *et al.* achieved crystalline layers with the parasitic reflection (103) when growing $Al_xIn_{1-x}N$ on quartz for growth temperatures below 300°C [35]. The improvement obtained in this study is attributed to the use of pure nitrogen as reactive gas, which reduces the growth rate and enhances the adatom diffusion during the growth even at low temperature thanks to the energy provided to the involved species in the plasma during the growth process.

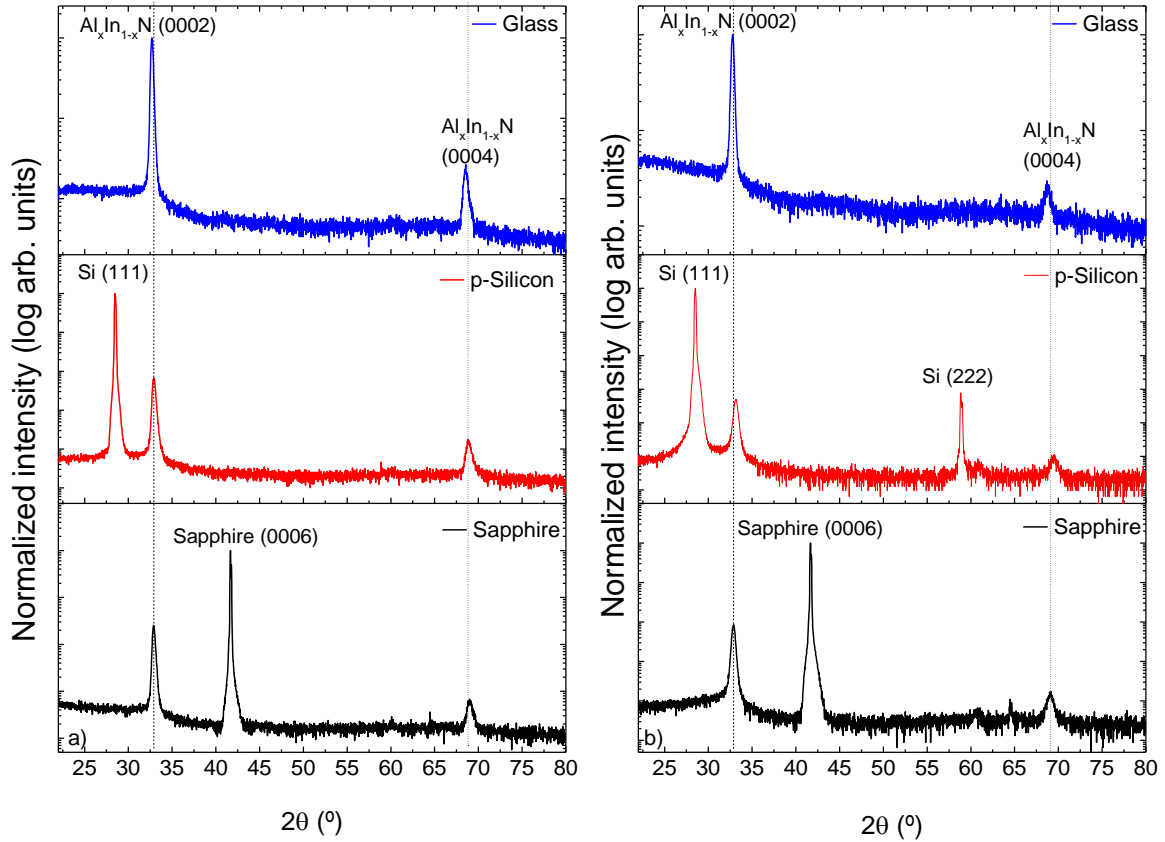

**Figure 1**. (color online) 2θ/ω scans of the $Al_xIn_{1-x}N$ layers grown on different substrates at a) room temperature and b) 100°C.

The Al content of the samples was estimated from HRXRD measurements assuming fully relaxed layers, considering the lattice parameters of InN ($c_{InN}$ = 5.703 Å) and AlN ($c_{AlN}$ = 4.982 Å) [44] and using the Vegard's law [45]. By this technique, the estimated Al content for the $Al_xIn_{1-x}N$ layers grown on sapphire and silicon at low temperature (RT and 100°C) is 37%. Previous studies show that the error with this assumption is below 4 % [27]. The layer composition of samples grown on sapphire and Si (111) at 300°C (see more details in [27]) was estimated by wavelength dispersive X-ray spectroscopy (WDX) measurements, revealing an aluminum composition of 36 and 37%, respectively. The similarity between the layer composition values obtained for the layers grown at low temperature by HRXRD measurements and the ones obtained by WDX measurements for the layers grown at 300ºC points to almost fully relaxed layers when growing at low temperatures. On the other hand, the estimation of the Al composition for the layers grown on glass substrates is highly affected by the lack of diffraction peak related to the substrate to be used as reference, obtaining a value of 33% (see Table 1).

The grain size (G) of the $Al_xIn_{1-x}N$ layer has been estimated by the Scherrer formula [46] $\left(G=\frac{0.9\lambda}{\Delta 2\theta}\cos\theta\right)$ which takes into account the X-ray wavelength $\lambda=1.54$ Å, the angle ($\theta$) and the FWHM of the (0002) $Al_xIn_{1-x}N$ diffraction peak ($\Delta 2\theta$). For all the studied substrates at low growth temperature (RT and 100°C) the estimated grain size lies in the range of 20 to 25 nm. Particularly, in the case of sapphire it increases from 25 to 53 nm when increasing the growth temperature from RT to 300°C. This increase of the grain size with the temperature is related with the enhancement of the adatom mobility that enables the growth and coalescence of the formed islands as reported by Afzal *et al.* for $Al_{0.17}In_{0.83}N$ layers grown on p-Si (111) [28].

Figure 2 shows a comparison between the rocking curve of the (0002) $Al_xIn_{1-x}N$ diffraction peak of the layers grown at RT and 300°C on sapphire (top) and on p-Si (111) (bottom) substrates. The observed reduction of the FWHM of the rocking curve with the substrate temperature is related to an increase in the adatom mobility due to the increase in substrate temperature which allows the enhancement of the crystal quality. As shown in Figure 2 and in Table 1, the FWHM of the rocking curve drops by a factor of ×4.6 and ×2 in the case of sapphire and silicon, respectively, when increasing the growth temperature from RT to 300°C. This reduction is not present when increasing the growth temperature from RT to 100°C, as the enhancement in adatom mobility is not enough to derive in a significant layer quality improvement. The $Al_xIn_{1-x}N$ layers grown on glass at low substrate temperature present the lowest FWHM of the rocking curve for each temperature among all the substrates used. This can be due to the amorphous nature of the glass, so that the $Al_xIn_{1-x}N$ layer does not have to compensate the lattice mismatch between layer and substrate. At the same time the amorphous nature of the substrate also hinders the pre-definition of preferential nucleation points for the growth of the subsequent layer, which usually determine the properties of the nitride layer, as it has been observed when growing InN on different buffer layers [47] (see Table 1).

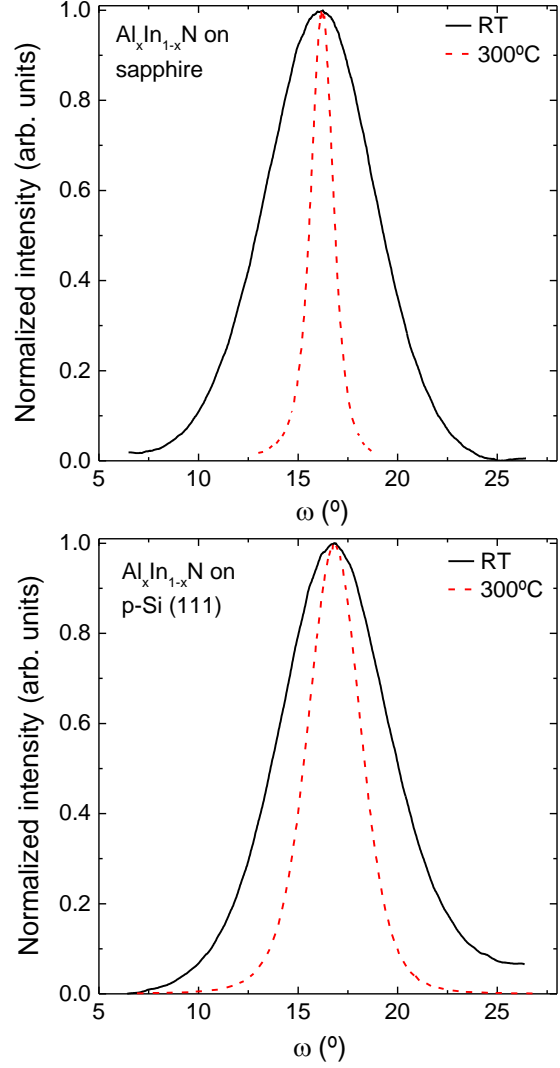

**Figure 2**. (color online) Rocking curve of the (0002) $Al_xIn_{1-x}N$ diffraction peak of the layers grown at different substrate temperatures, top: on sapphire substrate and bottom: on p-Si (111).

AFM and FESEM techniques were used to study the morphological properties of the layers. Higher substrate temperatures lead to an increase in the RMS surface roughness of the $Al_xIn_{1-x}N$ layers independently on the substrate (see Table 1). This can be due to the increase in the kinetic energy, and thus the mobility, of the sputtered atoms that occurs when increasing the growth temperature which leads to more compact layers with higher grain size as shown in the AFM images presented in Figure 3. These results are in good agreement with the ones obtained by HRXRD. This effect has been also observed by Afzal *et al.* in $Al_{0.17}In_{0.83}N$ layers grown on p-silicon (111) [28]. The effect of the substrate in the RMS surface roughness is studied for each growth temperature. At RT and 100°C all layers present similar RMS surface roughness independently of the substrate, with grain size and layer morphology similar for all substrates. This mean that at low growth temperatures, the substrate does not influence the growth of the subsequent AlInN layer. We have observed that there is a relationship between the grain size and the RMS surface roughness: the higher the grain size the higher the RMS. For example, the layer grown on silicon at RT presents the lowest grain size (19 nm) and the lowest RMS surface roughness (1.6 nm), and the layer grown on sapphire at 300°C presents the highest grain size (53 nm) and the highest RMS surface roughness (5.8 nm).

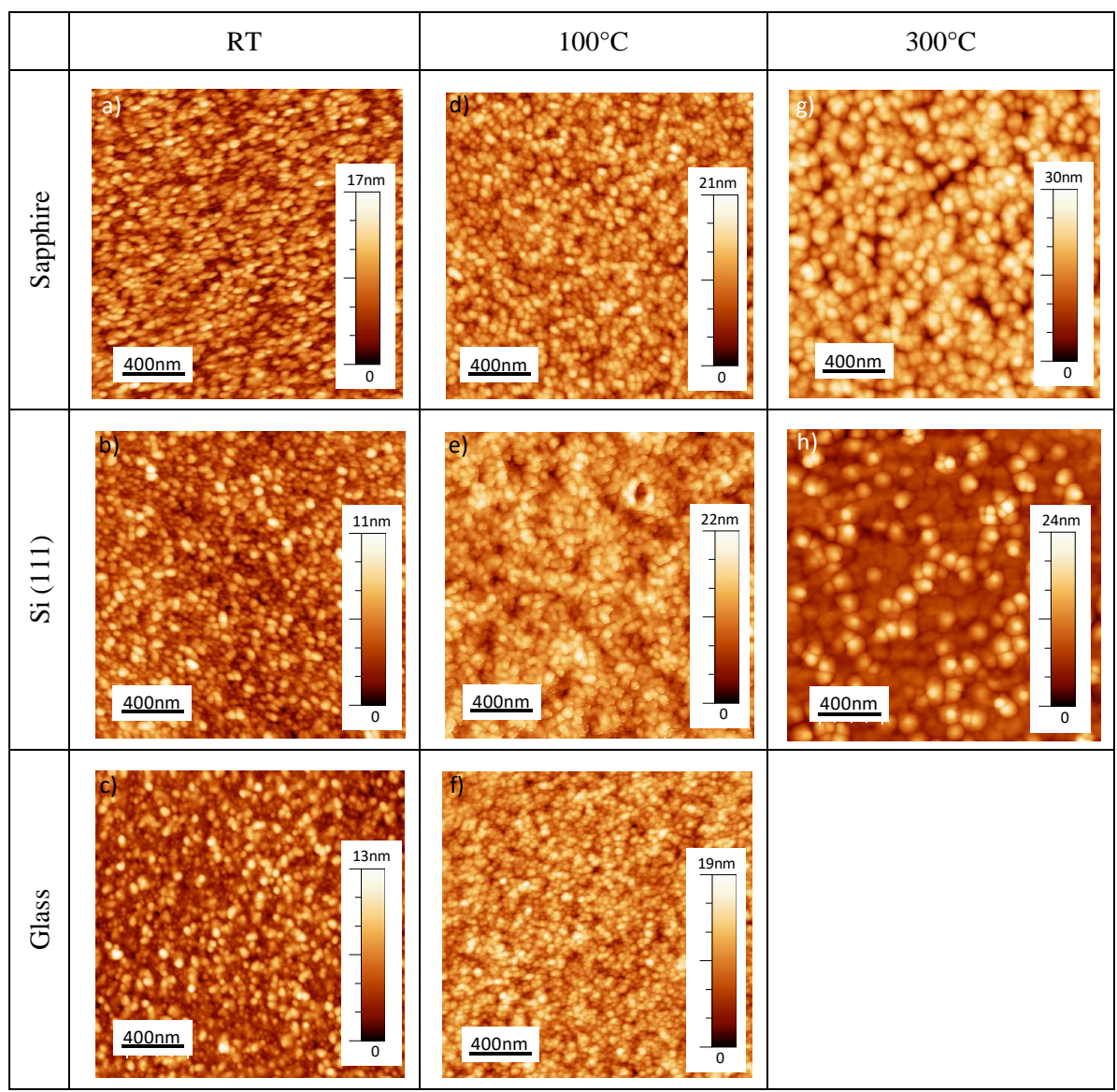

**Figure 3.** (color online) 2×2 μm² AFM images of $Al_{0.37}In_{0.63}N$ layer, the growth temperature and substrates are specified in each case.

The FESEM images of the layers grown on sapphire and Si (111) at RT and at 100°C are shown in Figure 4. AlInN layers grown on sapphire and Si (111) show a closely-packed columnar morphology for both temperatures (RT and 100°C). The improvement of this layer morphology was previously reported by our group: compact layers were achieved for $Al_xIn_{1-x}N$ grown on sapphire at 450°C (see [27]) and at 300°C for layers grown on p-Si (111) (see [26]). From these images, the film thickness is estimated as shown in Table 1.

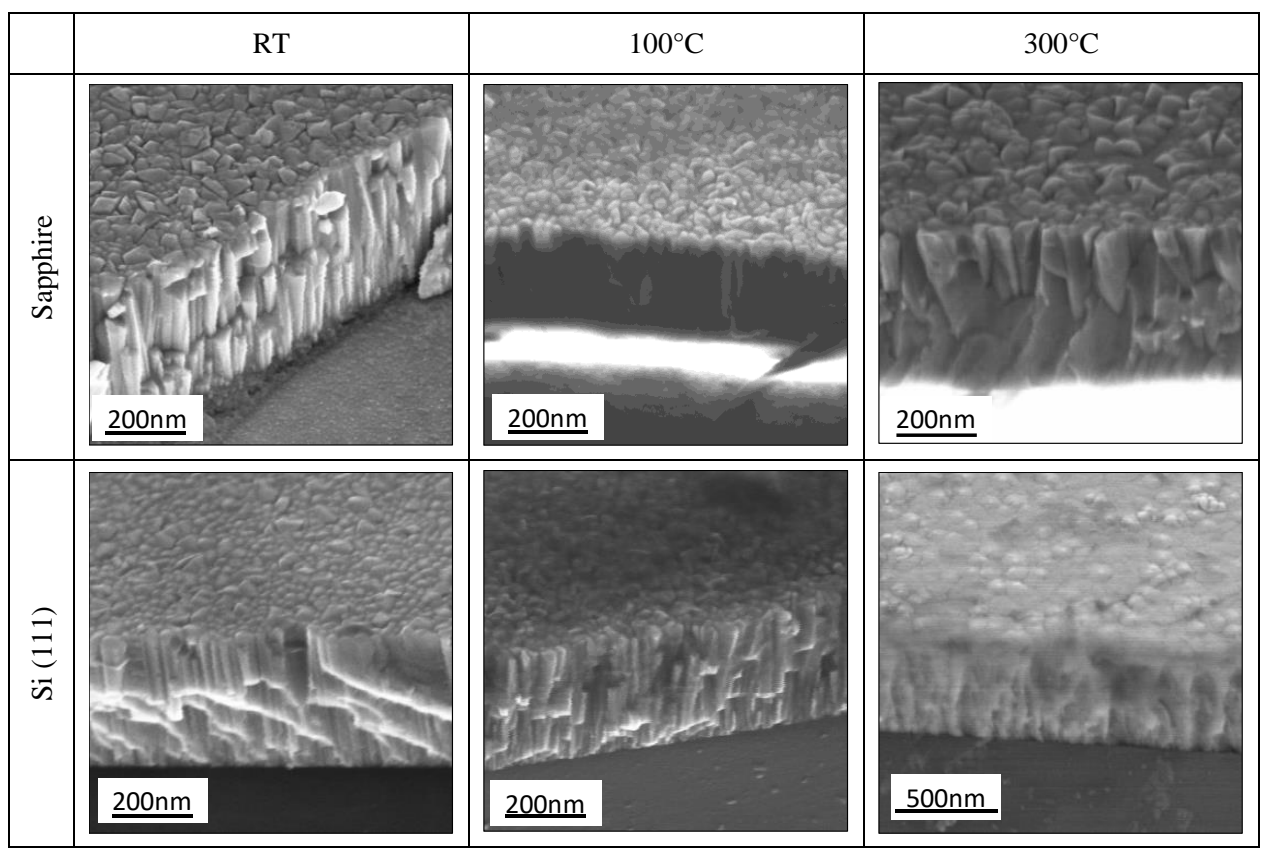

**Figure 4.** FESEM images of $Al_{0.37}In_{0.63}N$ layer, the growth conditions and substrate used in each case are indicated in the table.

*3.2. Optical properties*

The optical properties of the layers grown on sapphire and glass substrates were assessed by transmittance measurements. The inset of Figure 5 shows the optical transmittance of the $Al_xIn_{1-x}N$ layer grown on sapphire at RT and 300°C. Taking into account the relationship $\alpha(E) \propto -\ln(Tr)$, it is possible to estimate the absorption of the samples; this relation neglects optical scattering and reflection losses.

The square of the absorption coefficient has been calculated from a sigmoidal approximation ($\alpha = \alpha_0/1 + e^{\frac{E_0-E}{\Delta E}}$) of the absorption coefficient (shown in Figure 5). For the layers grown at RT a band gap energy of 2.34 and 2.37 eV has been calculated for sapphire and glass substrates, respectively. In the case of layers grown at 100°C the band gap energy has been estimated to be 2.44 and 2.46 eV for sapphire and glass substrates, respectively. Finally, for layers grown at 300°C on sapphire a band gap energy of 2.03 eV has been calculated (see Table 1), in accordance with previous results considering a bowing parameter of 5.2 eV [27].

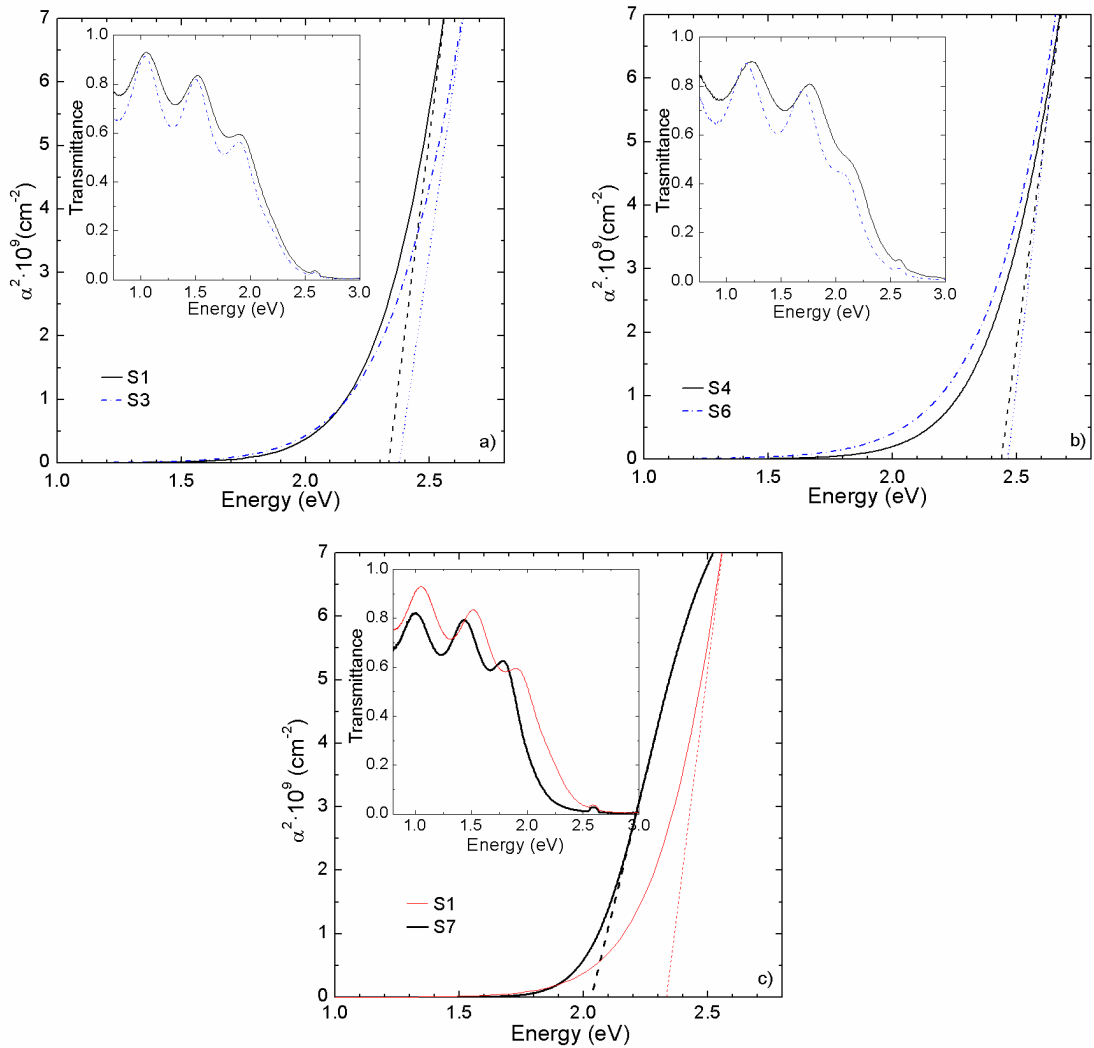

**Figure 5.** (color online) Square of the absorption coefficient as a function of the photon energy and inset RT transmittance of the $Al_xIn_{1-x}N$ layers grown on sapphire (black solid line) and glass (blue dashed line) at: a) RT, b) 100°C. c) comparison between RT and 300°C for sapphire substrate.

Hall effect measurements were carried out at RT to elucidate the origin in the difference in band gap energy of the AlxIn1-xN layers grown on sapphire at low (RT and 100°C) and high (300°C) substrate temperature. Layers deposited at 300ºC present values of resistivity ($1.1 \cdot 10^{-3}$ Ω·cm), mobility (~4 cm$^2$/V·s) and residual carrier concentration ($1.1 \cdot 10^{20}$ cm$^{-3}$) in good agreement with values reported in the literature [29]. On the contrary, the layers grown at low temperature present a high resistivity of 26 Ω·cm with low mobility (~0.3 cm$^2$/V·s) and residual carrier concentration ($6 \cdot 10^{17}$ cm$^{-3}$). The dramatic reduction of the layer mobility and the increase of its resistivity can be attributed to the highly columnar morphology of the samples deposited at RT, which increase the carrier scattering at the grain boundaries. At the same, this morphology type promotes the incorporation of oxygen impurities at grain boundaries, which has been related to an increase of the bandgap energy in InN films deposited by rf sputtering [48].

For layers grown on sapphire, there is a decrease of the absorption broadening (ΔE, obtained from the sigmoidal approximation of the absorption coefficient) from 288 and 154 meV when increasing the $T_s$ from RT and 300°C (see Table 1), which is attributed to a reduction in both alloy inhomogeneities and impurities incorporation at grain boundaries for layers grown at higher temperatures.

## 4. Conclusions

Al$_{0.37}$In$_{0.63}$N layers on sapphire, p-Si (111) and glass substrates were deposited by RF-sputtering at low (<300°C) growth temperature. The effect of the type of substrate and growth temperature on the properties of the layers was assessed. Polycrystalline Al$_{0.37}$In$_{0.63}$N layers oriented in the *c*-axis and with wurtzite structure have been achieved even at RT for all substrates. The FWHM of the rocking curve is reduced when increasing the growth temperature for all substrates due to the increase in the adatom mobility with the substrate temperature. The minimum FWHM of the rocking curve is obtained for layers grown on glass at RT and 100°C, which is attributed to the amorphous character of the substrate. The Al content estimated by HRXRD measurements for the layers grown at low temperature is similar to the one obtained by WDX measurements for Al$_x$In$_{1-x}$N layers grown at 300°C (37%) which points out to an almost fully relaxed state. The layers grown on sapphire and Si (111) at low temperatures show a closely-packed columnar like morphology with low RMS surface roughness (<3 nm). This morphology evolves to compact at growth temperatures of 300°C for $Al_{0.37}In_{0.63}N$ layers grown on Si (111). Nevertheless, a higher RMS surface roughness is obtained for layers grown at high temperature, probably due to an increase of the grain size with the temperature. The layers deposited on sapphire 300°C present a band gap energy of around 2.0 eV. However, there is a blue shift of the band gap energy to 2.4 eV when decreasing the growth temperature from 300°C to RT, which can be mainly attributed to the effect on oxygen incorporation at grain boundaries.

**Acknowledgements**

This work was partially supported by the projects TEC2014-60483-R, TEC2015-71127-C2-2-R, TEC2017-84378-R (Spanish Government); S2013/MIT 2790, P2018/EMT-4308 (Community of Madrid); CCG2015/EXP-014 (University of Alcalá). Co-financing from FEDER program is also acknowledged. The authors want to thank Dr. Eva Monroy from CEA-Grenoble (France) for technical and scientific support. The work of D. Montero was supported by the Spanish MINECO under contract BES – 2014 – 067585.

**References**

[1] J. Wu, W. Walukiewicz, K.M. Yu, J.W. Ager, E.E. Haller, H. Lu, W.J. Schaff, Y. Saito, Y. Nanishi, Unusual properties of the fundamental band gap of InN, Appl. Phys. Lett. 80 (2002) 3967–3969. doi:10.1063/1.1482786.

[2] W.M. Yim, E.J. Stofko, P.J. Zanzucchi, J.I. Pankove, M. Ettenberg, S.L. Gilbert, Epitaxially grown AlN and its optical band gap, J. Appl. Phys. 44 (1973) 292–296. doi:10.1063/1.1661876.

[3] K. Lorenz, N. Franco, E. Alves, I.M. Watson, R.W. Martin, K.P. O'Donnell, Anomalous ion channeling in AlInN/GaN bilayers: Determination of the strain state, Phys. Rev. Lett. 97 (2006) 1–4. doi:10.1103/PhysRevLett.97.085501.

[4] M. Gonschorek, J.-F. Carlin, E. Feltin, M. a. Py, N. Grandjean, High electron mobility lattice-matched AlInN∕GaN field-effect transistor heterostructures, Appl. Phys. Lett. 89 (2006) 062106. doi:10.1063/1.2335390.

[5] I. Saidi, H. Mejri, M. Baira, H. Maaref, Electronic and transport properties of AlInN/AlN/GaN high electron mobility transistors, Superlattices Microstruct. 84 (2015) 113–125. doi:10.1016/j.spmi.2015.04.036.

[6] J.F. Carlin, M. Ilegems, High-quality AlInN for high index contrast Bragg mirrors lattice matched to GaN, Appl. Phys. Lett. 83 (2003) 668–670. doi:10.1063/1.1596733.

[7] C. Berger, A. Dadgar, J. Bläsing, A. Lesnik, P. Veit, G. Schmidt, T. Hempel, J. Christen, A.

Krost, A. Strittmatter, Growth of AlInN/GaN distributed Bragg reflectors with improved interface quality, J. Cryst. Growth. 414 (2015) 105–109. doi:10.1016/j.jcrysgro.2014.09.008.

[8] A. Yamamoto, M.R. Islam, T.T. Kang, A. Hashimoto, Recent advances in InN-based solar cells: Status and challenges in InGaN and InAlN solar cells, Phys. Status Solidi. 7 (2010) 1309–1316. doi:10.1002/pssc.200983106.

[9] H.F. Liu, C.C. Tan, G.K. Dalapati, D.Z. Chi, Magnetron-sputter deposition of high-indium-content n-AlInN thin film on p-Si(001) substrate for photovoltaic applications, J. Appl. Phys. 112 (2012) 063114. doi:10.1063/1.4754319.

[10] H.F. Liu, S.B. Dolmanan, S. Tripathy, G.K. Dalapati, C.C. Tan, D.Z. Chi, Effects of AlN thickness on structural and transport properties of In-rich n-AlInN/AlN/p-Si(0 0 1) heterojunctions grown by magnetron sputtering, J. Phys. D. Appl. Phys. 46 (2013) 095106. doi:10.1088/0022-3727/46/9/095106.

[11] H. Kim-Chauveau, P. de Mierry, J.-M. Chauveau, J.-Y. Duboz, The influence of various MOCVD parameters on the growth of Al1−xInxN ternary alloy on GaN templates, J. Cryst. Growth. 316 (2011) 30–36. doi:10.1016/j.jcrysgro.2010.12.040.

[12] D.A. Neumayer, J.G. Ekerdt, Growth of Group III Nitrides. A Review of Precursors and Techniques, Chem. Mater. 8 (1996) 9–25. doi:10.1021/cm950108r.

[13] C. Hums, J. Bläsing, A. Dadgar, A. Diez, T. Hempel, J. Christen, A. Krost, K. Lorenz, E. Alves, Metal-organic vapor phase epitaxy and properties of AlInN in the whole compositional range, Appl. Phys. Lett. 90 (2007) 022105. doi:10.1063/1.2424649.

[14] M. Ferhat, F. Bechstedt, First-principles calculations of gap bowing in $In_xGa_{1-x}N$ and $In_xAl_{1-x}N$ alloys: Relation to structural and thermodynamic properties, Phys. Rev. B. 65 (2002) 075213. doi:10.1103/PhysRevB.65.075213.

[15] S. Yamaguchi, M. Kariya, S. Nitta, H. Kato, T. Takeuchi, C. Wetzel, H. Amano, I. Akasaki, Structural and optical properties of AlInN and AlGaInN on GaN grown by metalorganic vapor phase epitaxy, J. Cryst. Growth. 195 (1998) 309–313. doi:10.1016/S0022-0248(98)00629-0.

[16] T. Aschenbrenner, H. Dartsch, C. Kruse, M. Anastasescu, M. Stoica, M. Gartner, A. Pretorius, A. Rosenauer, T. Wagner, D. Hommel, Optical and structural characterization of AlInN layers for optoelectronic applications, J. Appl. Phys. 108 (2010) 063533. doi:10.1063/1.3467964.

[17] L. Yun, T. Wei, J. Yan, Z. Liu, J. Wang, J. Li, MOCVD epitaxy of InAlN on different templates, J. Semicond. 32 (2011) 093001. doi:10.1088/1674-4926/32/9/093001.

[18] T.T. Kang, M. Yamamoto, M. Tanaka, A. Hashimoto, A. Yamamoto, Effect of gas flow on the growth of In-rich AlInN films by metal-organic chemical vapor deposition, J. Appl. Phys. 106 (2009) 7–10. doi:10.1063/1.3212969.

[19] W.C. Chen, Y.H. Wu, C.Y. Peng, C.N. Hsiao, L. Chang, Effect of In/Al ratios on structural and optical properties of InAlN films grown on Si(100) by RF-MOMBE, Nanoscale Res. Lett. 9 (2014) 1–7. doi:10.1186/1556-276X-9-204.

[20] Y.H. Wu, Y.Y. Wong, W.C. Chen, D.S. Tsai, C.Y. Peng, J.S. Tian, L. Chang, E. Yi Chang, Indium-rich InAlN films on GaN/sapphire by molecular beam epitaxy, Mater. Res. Express. 1 (2014) 015904. doi:10.1088/2053-1591/1/1/015904.

[21] H. Naoi, K. Fujiwara, S. Takado, M. Kurouchi, D. Muto, T. Araki, H. Na, Y. Nanishi, Growth of in-rich $in_xAl_{1-x}N$ films on (0001) sapphire by RF-MBE and their properties, J. Electron. Mater. 36 (2007) 1313–1319. doi:10.1007/s11664-007-0195-4.

[22] J. Kamimura, T. Kouno, S. Ishizawa, A. Kikuchi, K. Kishino, Growth of high-In-content InAlN nanocolumns on Si (1 1 1) by RF-plasma-assisted molecular-beam epitaxy, J. Cryst. Growth. 300 (2007) 160–163. doi:10.1016/j.jcrysgro.2006.11.029.

[23] W. Terashima, S.B. Che, Y. Ishitani, A. Yoshikawa, Growth and characterization of AlInN ternary alloys in whole composition range and fabrication of InN/AlInN multiple quantum wells by RF molecular beam epitaxy, Jpn. J. Appl. Phys. 45 (2006) L539–L542.

doi:10.1143/JJAP.45.L539.

[24] Q. Guo, T. Tanaka, M. Nishio, H. Ogawa, Structural and Optical Properties of AlInN Films Grown on Sapphire Substrates, Jpn. J. Appl. Phys. 47 (2008) 612. doi:10.1143/JJAP.47.612.

[25] Q. Guo, K. Yahata, T. Tanaka, M. Nishio, H. Ogawa, Growth and characterization of reactive sputtered AlInN films, Phys. Status Solidi. 0 (2003) 2533–2536. doi:10.1002/pssc.200303378.

[26] A. Núñez-Cascajero, L. Monteagudo-Lerma, S. Valdueza-Felip, C. Navío, E. Monroy, M. González-Herráez, F.B. Naranjo, Study of high In-content AlInN deposition on p-Si ( 111 ) by RF-sputtering, Jpn. J. Appl. Phys. 55 (2016) 05FB07. doi:10.7567/JJAP.55.05FB07.

[27] A. Núñez-Cascajero, S. Valdueza-Felip, L. Monteagudo-Lerma, E. Monroy, E. Taylor, R. Martin, M. Gonzalez-Herraez, F.. Naranjo, In-rich AlxIn1-xN grown by RF-sputtering on sapphire: from closely-packed columnar to high-surface quality compact layers, J. Phys. D. Appl. Phys. 50 (2016) 1–9. doi:10.1088/1361-6463/aa53d5.

[28] N. Afzal, M. Devarajan, K. Ibrahim, Influence of substrate temperature on the growth and properties of reactively sputtered In-rich InAlN films, J. Mater. Sci. Mater. Electron. 27 (2016) 4281–4289. doi:10.1007/s10854-016-4294-y.

[29] N. Afzal, M. Devarajan, K. Ibrahim, A comparative study on the growth of InAlN films on different substrates, Mater. Sci. Semicond. Process. 51 (2016) 8–14. doi:10.1016/j.mssp.2016.04.004.

[30] T.-S. Yeh, J.-M. Wu, W.-H. Lan, The effect of AlN buffer layer on properties of AlxIn1−xN films on glass substrates, Thin Solid Films. 517 (2009) 3204–3207. doi:10.1016/j.tsf.2008.10.101.

[31] Q. Han, C. Duan, G. Du, W. Shi, L. Ji, Magnetron sputter epitaxy and characterization of wurtzite AlInN on Si(111) substrates, J. Electron. Mater. 39 (2010) 489–493. doi:10.1007/s11664-010-1112-9.

[32] C. Besleaga, A.C. Galca, C.F. Miclea, I. Mercioniu, M. Enculescu, G.E. Stan, A.O. Mateescu, V. Dumitru, S. Costea, Physical properties of AlxIn1-xN thin film alloys sputtered at low temperature, J. Appl. Phys. 116 (2014). doi:10.1063/1.4898565.

[33] C.J. Dong, M. Xu, Q.Y. Chen, F.S. Liu, H.P. Zhou, Y. Wei, H.X. Ji, Growth of well-oriented AlxIn1-xN films by sputtering at low temperature, J. Alloys Compd. 479 (2009) 812–815. doi:10.1016/j.jallcom.2009.01.075.

[34] L.F. Jiang, W.Z. Shen, Q.X. Guo, Temperature dependence of the optical properties of AlInN, J. Appl. Phys. 106 (2009) 013515. doi:10.1063/1.3160299.

[35] H. He, Y. Cao, R. Fu, W. Guo, Z. Huang, M. Wang, C. Huang, J. Huang, H. Wang, Band gap energy and bowing parameter of In-rich InAlN films grown by magnetron sputtering, Appl. Surf. Sci. 256 (2010) 1812–1816. doi:10.1016/j.apsusc.2009.10.012.

[36] H. He, Y. Cao, R. Fu, H. Wang, J. Huang, C. Huang, M. Wang, Z. Deng, Structure and optical properties of InN and InAlN films grown by rf magnetron sputtering, J. Mater. Sci. Mater. Electron. 21 (2010) 676–681. doi:10.1007/s10854-009-9976-2.

[37] M. Lü, C. Dong, Y. Wang, Proposal and achievement of a relatively Al-rich interlayer for In-rich AlxIn1−xN films deposition, J. Wuhan Univ. Technol. Sci. Ed. 28 (2013) 868–875. doi:10.1007/s11595-013-0784-4.

[38] K. Kubota, Y. Kobayashi, K. Fujimoto, Preparation and properties of III-V nitride thin films, J. Appl. Phys. 66 (1989) 2984–2988. doi:10.1063/1.344181.

[39] T. Peng, J. Piprek, G. Qiu, J.O. Olowolafe, K.M. Unruh, C.P. Swann, E.F. Schubert, Band gap bowing and refractive index spectra of polycrystalline AlxIn1-xN films deposited by sputtering, Appl. Phys. Lett. 71 (1997) 2439–2441. doi:10.1063/1.120112.

[40] Q.X. Guo, Y. Okazaki, Y. Kume, T. Tanaka, M. Nishio, H. Ogawa, Reactive sputter deposition of AlInN thin films, J. Cryst. Growth. 300 (2007) 151–154. doi:10.1016/j.jcrysgro.2006.11.007.

[41] Q. Guo, N. Shingai, Y. Mitsuishi, M. Nishio, H. Ogawa, Effects of nitrogen/argon ratio on composition and structure of InN films prepared by r.f. magnetron sputtering, Thin Solid Films. 343–344 (1999) 524–527. doi:10.1016/S0040-6090(98)01671-X.

[42] S. Valdueza-Felip, F.B. Naranjo, M. González-Herráez, L. Lahourcade, E. Monroy, S. Fernández, Influence of deposition conditions on nanocrystalline InN layers synthesized on Si(1 1 1) and GaN templates by RF sputtering, J. Cryst. Growth. 312 (2010) 2689–2694. doi:10.1016/j.jcrysgro.2010.05.036.

[43] I. Horcas, R. Fernández, J.M. Gómez-Rodríguez, J. Colchero, J. Gómez-Herrero, a. M. Baro, WSXM: A software for scanning probe microscopy and a tool for nanotechnology, Rev. Sci. Instrum. 78 (2007) 013705. doi:10.1063/1.2432410.

[44] I. Vurgaftman, J.R. Meyer, Band parameters for nitrogen-containing semiconductors, J. Appl. Phys. 94 (2003) 3675–3696. doi:10.1063/1.1600519.

[45] L. Vegard, Die Konstitution der Mischkristalle und die Raumfüllung der Atome, - Zeitschrift Für Phys. 5 (1921) 17–26. doi:10.1007/BF01349680.

[46] A.L. Patterson, The Scherrer Formula for X-Ray Particle Size Determination, Phys. Rev. 56 (1939) 978–982. doi:10.1103/PhysRev.56.978.

[47] L. Monteagudo-Lerma, S. Valdueza-Felip, A. Núñez-Cascajero, A. Ruiz, M. González-Herráez, E. Monroy, F.B.B. Naranjo, Morphology and arrangement of InN nanocolumns deposited by radio-frequency sputtering: Effect of the buffer layer, J. Cryst. Growth. 434 (2016) 13–18. doi:10.1016/j.jcrysgro.2015.10.016.

[48] Motlan, E.M. Goldys, T.L. Tansley, Optical and electrical properties of InN grown by radio-frequency reactive sputtering, J. Cryst. Growth. 241 (2002) 165–170.

**Figure captions:**

Figure 1. (color online) 2θ/ω scans of the $Al_xIn_{1-x}N$ layers grown on different substrates at a) room temperature and b) 100°C.

Figure 2. (color online) Rocking curve of the (0002) $Al_xIn_{1-x}N$ diffraction peak of the layers grown at different substrate temperatures, top: on sapphire substrate and bottom: on p-Si (111).

Figure 3. (color online) 2×2 µm2 AFM images of $Al_{0.37}In_{0.63}N$ layer, the growth temperature and substres are specified in each case.

Figure 4. FESEM images of $Al_{0.37}In_{0.63}N$ laye taken with a tilt angle of 45°, the growth conditions and substrate used in each case is indicated in the table.

Figure 5. (color online) Square of the absorption coefficient as a function of the photon energy and inset RT transmittance of the $Al_xIn_{1-x}N$ layers grown on sapphire and glass at: a) RT and b) 100°C.